\documentclass[aps,twocolumn,nofootinbib,superscriptaddress,amsfonts,floatfix
]{revtex4-1} 
\usepackage{graphicx,amsmath,amssymb,amstext}
\usepackage{amssymb,amsbsy,amsfonts,amsthm,color}

\usepackage{mathtools,nccmath}

\usepackage{epsfig}
\usepackage{graphicx}
\usepackage{subfigure}
\usepackage[dvipsnames]{xcolor}
\definecolor{linkcolor}{rgb}{0.0,0.3,0.5}
\usepackage[unicode, colorlinks=true, linkcolor=linkcolor, citecolor=linkcolor, 
filecolor=linkcolor,urlcolor=linkcolor, pdfusetitle]{hyperref}
\usepackage{cancel}
\usepackage{balance}
\usepackage[capitalise]{cleveref}
\usepackage{xspace}
\usepackage{enumerate}
\usepackage{ulem}
\usepackage{mdframed}
\usepackage{booktabs}
\usepackage{makecell}
\usepackage{enumitem,tabularx,multirow}
\usepackage{aas_macros}
\usepackage{colortbl}
\usepackage{orcidlink}

\AtBeginDocument{%
  \heavyrulewidth=.08em
  \lightrulewidth=.05em
  \cmidrulewidth=.03em
  \belowrulesep=.65ex
  \belowbottomsep=0pt
  \aboverulesep=.4ex
  \abovetopsep=0pt
  \cmidrulesep=\doublerulesep
  \cmidrulekern=.5em
  \defaultaddspace=.5em
}

\graphicspath{{Figures/}}

\usepackage{color}

\begin{document}
\title{Self-consistent thermodynamical treatment for quark matter in quasi-particle model at finite temperature
}

\author{Suman Pal,\orcidlink{0009-0000-5944-4261}}
\email{sumanvecc@gmail.com}

\affiliation{Physics Group, Variable Energy Cyclotron Centre, 1/AF Bidhan Nagar, Kolkata 700064, India}
\affiliation{Homi Bhabha National Institute, Training School Complex, Anushakti Nagar, Mumbai 400085, India}
\author{Gargi Chaudhuri,\orcidlink{0000-0002-8913-0658}}
\email{gargi@vecc.gov.in}
\affiliation{Physics Group, Variable Energy Cyclotron Centre, 1/AF Bidhan Nagar, Kolkata 700064, India}
\affiliation{Homi Bhabha National Institute, Training School Complex, Anushakti Nagar, Mumbai 400085, India}

\begin{abstract}

In this work, we have studied the medium effects in strange quark matter in the framework of a grand-canonical ensemble using the phenomenological quasi-particle model. This model is studied with proper self-consistent thermodynamical treatment by incorporating chemical potential-dependent quark mass. We have also included the vector interaction in a self-consistent way. The main aim of this work is to explore the proper thermodynamic treatment in addressing the medium effects at both zero and finite temperatures. In the case of the finite temperature, we explore the study of self-consistent thermodynamics in the isothermal as well as the isentropic processes. The effect of finite temperature and lepton fraction have been studied on the equation of state, speed of sound, and particle fraction. The  $M-R$ and  $M-\Lambda$ diagrams are found to be consistent with the observational constraints.

\end{abstract}
\maketitle


\section{Introduction} 
\label{sect:intro} 

The exploration of the equation of state for strongly interacting dense matter within neutron stars (NSs) \cite{Antoniadis:2013pzd, Fonseca:2021wxt, Riley:2021pdl, Miller:2021qha, Riley:2019yda, LIGOScientific:2017vwq, Radice:2017lry} is a subject of significant contemporary interest \cite{Baym:2017whm, Lovato:2022vgq}. It is a common belief that quark matter may exist within neutron stars (NS)\cite{Annala:2019puf, Annala:2019puf}. However, first-principles methods are not suitable for describing quark matter at the densities found in the core of NSs. 
This limitation stems from the sign problem encountered in lattice Monte Carlo simulations at non-zero chemical potentials \cite{deForcrand:2009zkb} and the fact that perturbative QCD remains effective only at much higher densities \cite{Kurkela:2009gj}.
Numerous efforts have been undertaken to include nonperturbative effects in increasingly advanced models given the fact  that perturbative QCD is insufficient for determining the equation of state (EoS) of quark matter.

Strange quark matter(SQM) plays an important role in many interesting fields for example in hot and dense matter in heavy ion collision, the structure of compact stars, etc.  Ever since W. H. Witten suggested \cite{Farhi:1984qu,torres2013quark, Ferrer:2015}that the SQM would be absolutely stable even at absolute zero temperature, there has been a lot of interest in studying it. 
At extremely high densities, the concept of asymptotic freedom in QCD implies the potential occurrence of a first-order phase transition from hadronic matter to quark matter. This leads to the possible formation of hybrid stars(HS). Typically, the transition from nuclear matter to quark matter is expected to take place in the density range of nearly about 2 to 4 times $\rho_0$ \cite{Bhattacharyya:2009fg,Han:2019bub,Ferreira:2020evu,Sen:2022lig,suman2023a},  where $\rho_0$ is the nuclear saturation density.\\
In recent studies, researchers have extensively employed phenomenological quark models like the MIT bag model \cite{chodes1974,glendenning2012compact, Sen:2021cgl, Sen:2022lig,suman2023a, Podder:2023dey} and quark mass model or Quasi particle model\cite{Zhang:2021qhl,pen:23prc_qmdd, Ma:prd23sep,peng2001a,wen2005a, Chu_2014, Benvenuto95} to investigate the thermodynamic properties of strange quark matter, quark stars, and hybrid stars. These models typically account for all interactions among quarks through  medium dependent bag pressure or an equivalent quark mass. Recently in Ref.~\cite{Lugones},  the phenomenological density-dependent quark mass model has been revisited and the thermodynamical inconsistency has been resolved within the canonical ensemble formulation. In Ref.~\cite{pal:23prd2_bag} the authors have studied the MIT bag model and showed that if the medium effects are incorporated through  density-dependent bag pressure in the grand canonical ensemble, then the Euler relation is violated. If the Euler relation is used then the minimum of energy per baryon does not occur at zero pressure. To overcome this inconsistency, they have proposed the medium effect of the strange quark matter in the form of chemical potential-dependent bag pressure in the grand canonical ensemble. In the grand-canonical ensemble, the proper variable is the chemical potential whereas in the canonical ensemble, density is the appropriate intensive quantity. It is shown that in the grand-canonical ensemble, the density gets modified whereas chemical potential is modified in the canonical ensemble due to the medium effects.\\
In the past, the self-consistent thermodynamic treatment for the quasi-particle model were done on \cite{Chu:2021aaz, Schertler:1996tq,Chu:2022ofc,Zhang:2021qhl,pen:23prc_qmdd,Ma:prd23sep,Chu:2019ipr}. In this approach, a counterterm is introduced externally in the thermodynamic potential such that the extra term in the density, which arises due to the chemical potential-dependent mass, is offset by the derivative of the counterterm. In this work, our main aim is to propose a method to introduce the medium effect in the quasi-particle model respecting thermodynamic self-consistency without adding any  extra term .  We  use the standard prescription of starting  from the partition function of free Fermi gas, with quark mass modified to account for the medium effect, adding  bag pressure for the confinement, and then derive the thermodynamical potential from the partition function using the standard definition \cite{Kapusta:2023eix}.  We have shown that there is  absolutely no need to include 
any counter term for validating the self-consistency.  We emphasize that in our formalism there is no need to modify the thermodynamical potential externallyand that all quantities are derived following standard definitions.

In this work, we follow the prescription as in Ref.~\cite{pal:23prd2_bag}.  If we take the medium effect through chemical potential dependent mass, then the natural choice is the use of a grand-canonical ensemble, where the thermodynamic potential is an explicit function of $(\mu, T)$ and the density can be derived. In the previous version of the quasi-particle model \cite{Chu:2021aaz, Schertler:1996tq,Zhang:2021qhl,pen:23prc_qmdd, Ma:prd23sep, Chu_2014}, the density is unchanged due to adding of external term , but here in our formalism density is modified due to chemical potential-dependent quark mass.
To test our result we study both cold and hot quark matter emphasizing the proto-quark star (PQS) environment. In the case of the finite temperature equation of state, we have considered two scenarios: the isothermal process, where the temperature remains constant, and the pressure can be expressed as 
\begin{equation}\label{eq:f_min}
    P=-\left(\frac{\partial F}{\partial V}\right)_{T,N}=\rho^2 \frac{\partial}{\partial \rho}\left(\frac{f}{\rho}\right)_T,
\end{equation}
where \(f\) represents the free energy density.
According to Eq.~\eqref{eq:f_min}, the minimum of \(\frac{f}{\rho}\) should occur when the pressure is zero.

In contrast, for an isentropic process, where the entropy density remains constant, the pressure is given by 
\begin{equation}\label{eq:energy_min}
    P=-\left(\frac{\partial U}{\partial V}\right)_{S,N}=\rho^2 \frac{\partial}{\partial \rho}\left(\frac{\varepsilon}{\rho}\right)_{\sigma},
\end{equation}
where \(\varepsilon\) represents the energy density and $\sigma$ reprents entropy density per baryon density. In this case, the minimum of \(\frac{\varepsilon}{\rho}\) also occurs at zero pressure. 
In this work, we have studied both neutrino-trapped matter and neutrino-free matter which will be discussed in detail in  section \ref{sec:results}. 

This paper is organized as follows. In Sec.~\ref{sec:formalism} we give the detailed formalism of the equation of state within the framework of the grand canonical ensemble. In Sec.~\ref{sec:results} we show the numerical results for different observables. Finally, we summarise in Sec.~\ref{sec:conclusion}.
\section{Formalism}\label{sec:formalism}

Quark stars are supposed to be composed entirely of quark matter along with a minor fraction of electrons. The description of the quark matter in the grand-canonical ensemble (GC) deals with the grand partition function of free Fermi gas ${Z(\mu,V,T)}$, which depends on the chemical potential ($\mu$), volume ($V$) and temperature ($T$). All thermodynamic quantities can in turn be deduced from the grand-canonical potential $\bar{\Omega}(\mu, V, T)$. The thermodynamic potential density is defined as
\begin{equation}\label{def_omega}
   {\Omega(\mu,T)}=\frac{-T \log(Z(\mu,V,T))}{V}
\end{equation}
In this  ensemble, pressure is defined as
\begin{equation}\label{eq:pressure_def1}
P = -\left(\frac{\partial \bar{\Omega}}{\partial V}\right)_{T,\mu}
\end{equation}

where $\bar{\Omega}$ is the grand potential, 
$\Omega=\frac{\bar{\Omega}}{V}$ is grand-potential per unit volume and $N=\rho V$. We can write Eq.(\ref{eq:pressure_def1}) as
\begin{equation}\label{eq:pressure_def2}
P=-\left[\frac{\partial(\Omega V)}{\partial(\frac{N}{\rho})}\right]_{T,\mu}=-\Omega+\rho\left[\frac{\partial \Omega}{\partial \rho}\right]_{T,\mu}
\end{equation}  
$\Omega$ is independent of $\rho$, therefore second term of the Eq.~ \eqref{eq:pressure_def2} vanishes.
\begin{equation} \label{eq:pressure_def_final}
    P=-\Omega
\end{equation} 
The energy density is obtained by using the Euler's relation as 
\begin{equation}\label{eq:euler_rel}
    \varepsilon=-P+\sum_{i=u,d,s,e} \rho_i\mu_i+Ts
\end{equation} 
Here $\rho_i$ is the desnity of the $i_{\text{th}}$ particle and $\mu_i$ is the chemical potential of the $i_{\text{th}}$ particle and $s$ is the entropy density.
Free energy density is given by
\begin{equation}
    f=\varepsilon-Ts
\end{equation}

\begin{figure*}[!htbp]
    \centering
    \includegraphics[width=0.45\textwidth]{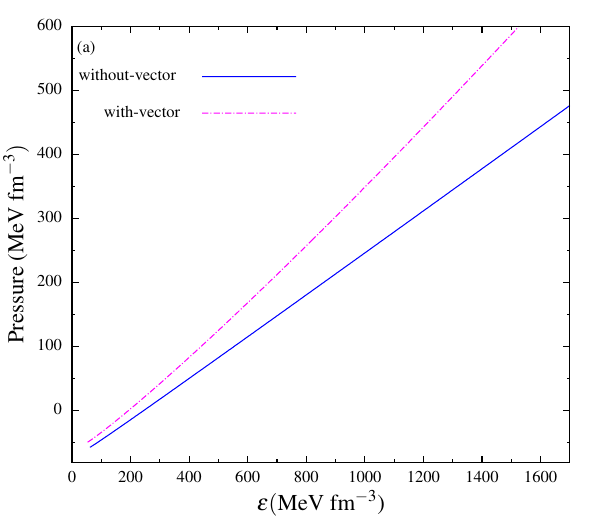}
    \includegraphics[width=0.45\textwidth]{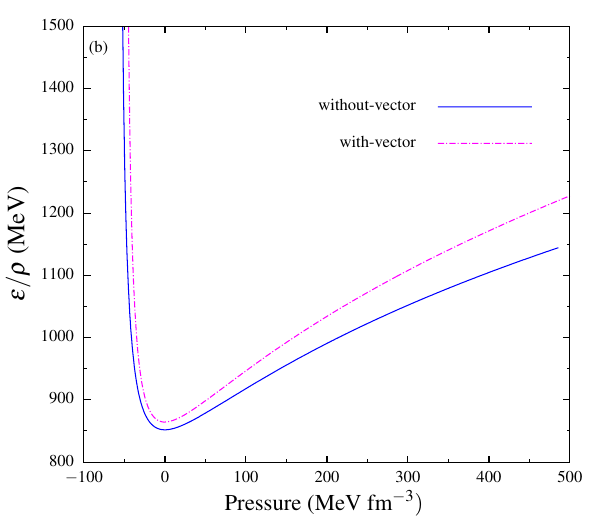}
    \includegraphics[width=0.45\textwidth]{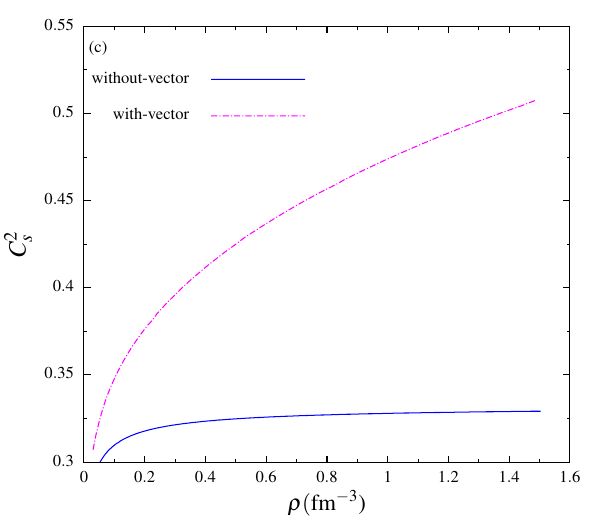}
    \includegraphics[width=0.45\textwidth]{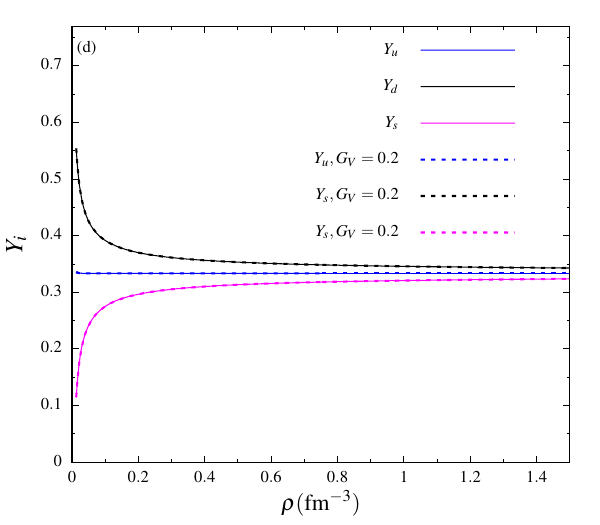}
    \caption{Plots for(a) pressure vs. energy density (b)$\frac{\varepsilon}{\rho}$ vs. pressure (c) speed of sound vs. density(d) and particle fraction with density at zero temperature without and with vector interactions. Model parameters are as follows: For the case without vector interactions,  $g_0 = 1.0$, $\alpha_{\mu} = 20.0$, and $B_0 = 58 \text{ MeV fm}^{-3}$. For the case with vector interactions, the parameters are $g_0 = 1.0$, $\alpha_{\mu} = 20.0$, $B_0 = 50 \text{ MeV fm}^{-3}$, and $G_V = 0.2 \text{ fm}^{-2}$.  }
    \label{fig:zero_temp_eos}
   \end{figure*} 

\subsubsection*{\textbf{ Medium effect at finite tempearture}} 
 In the quasi-particle model, the medium effect is taken through chemical potential-dependent quark mass while the bag pressure is taken to be constant ($B_0$). 
In the dense system, quarks interact with other quarks to create an effective mass, which makes them act as quasiparticles.
In the dense system, quarks interact with other quarks to create an effective mass, which represents the interactions between the quarks. In the hard dense loop approximation, an effective quark propagator is generated by resumming one-loop self-energy diagrams is used to determine the zero momentum limit of the dispersion relations, which leads to the effective quark masses as given by  \cite{Schertler:1996tq}
\begin{equation}\label{eq:quasi_mass}
m^*_i=\frac{m_{i0}}{2}+\sqrt{\frac{m_{i0}^2}{4}+\frac{g_i^2}{6\pi^2}\mu_i^2}
\end{equation} 
A previous study of the chemical potential dependent quark mass  states that \cite{Zhang:2021qhl,pen:23prc_qmdd,Ma:prd23sep}
in essence, the running coupling constant ideally shouldn't remain constant, and the effective quark mass is expected to decrease as $\mu_i$ increases in the high baryon density region. This adjustment is necessary to align with the QCD characteristics and ensure the reinstatement of chiral symmetry.
We have used the expression for $g_i$ as follows \cite{Zhang:2021qhl}
\begin{equation}\label{eq:quasi_mass_coupling}
    g_i=g_{0}e^{-\alpha_{\mu} \frac{\mu_i}{\mu_0}}
\end{equation} 
Here $\alpha_{\mu}$ is the parameter determining the $\mu$ dependent effective running coupling constant and the value of $\alpha_{\mu}$ should be greater than zero for the restoration of the chiral symmetry. 
The model parameters  $g_0,\alpha_{\mu}, B_0$ are adjusted by the stability criteria as mentioned in the introduction section \cite{Dexheimer:2013eua,pal:23prd2_bag}. 
In cold quark matter, the stability criterion requires that for three-flavor quark matter, the energy per baryon $(\frac{\varepsilon}{\rho})$ must be less than 930 MeV, while for two-flavor quark matter, it should exceed 930 MeV. At finite temperature, we consider the free energy density, where for three-flavor quark matter, $\frac{f}{\rho}$ must also be less than 930 MeV. Additionally, the parameter choices should ensure that the stability condition is satisfied at zero temperature as well.

In this study, we have not incorporated any counterterms in the thermodynamic potential as mentioned earlier. We write the thermodynamic potential for fermi gas of quarks with medium effect taken through chemical potential dependent quark mass, with constant bag pressure. 
The partition function of quark matter\cite{Kapusta:2023eix} for  the $i_{th}$ particle can be written as
\begin{equation} 
\begin{aligned}
     Z_i =& \exp\left[\frac{\gamma_i V}{2\pi^2} \int_0^{\infty} k^2 \, dk \left( \frac{E_i}{T} + \log\left(1 + e^{-\frac{E_i - \mu_i}{T}}\right) \right.\right.\\
      & \left.\left. + \log\left(1 + e^{-\frac{E_i +\mu_i}{T}}\right) \right) \right] \\
\end{aligned}
\end{equation} 
Here $\gamma_i$ is the particle's degeneracy(6 for quarks, 2 for electrons, and 1 for neutrinos) and Here $E_i(k)=\sqrt{k^2+(m_i^*)^2}$.
In quark matter, we have considered the components to be the three quarks  u, d and s as well as the leptons e and $\nu_e$. 
The thermodynamic potential of the quark matter with bag pressure $(B_0)$ (using Eq.~\eqref{def_omega}) is given by 
\begin{equation}
\begin{aligned}
    \Omega_{quasi}(T,\mu)=&\sum_{i=u,d,s}[-\frac{\gamma_i}{2\pi^2}\int_0^{\infty}k^2 \, dk \,\sqrt{k^2+(m_i^{*})^2}+\\&\frac{\gamma_i}{2\pi^2}\int_0^{\infty}k^2 \, dk \, T\left[\log(1-f_i^+)+\log(1-f_i^-)\right]] \\ &
    +\sum_{l=e,\nu_e}\Omega_l+B_0 \\
\end{aligned}
\end{equation}
The first term represents the vacuum term; we can drop the vacuum contribution \cite{Negele:1986bp}. Here $\Omega_l$ represents the electron and neutrino contributions. 
The thermodynamic potential  can be  then written as,
\begin{equation}\label{eq:quasi_pressure}
\begin{aligned}
    \Omega_{quasi}=&-\sum_{f=u,d,s,}\frac{1}{3}\frac{\gamma_i}{2\pi^2}\int_0^{\infty}\frac{k^4}{\sqrt{k^2+(m_f^*)^2}}[f_i^++f_i^{-}] dk+\\ &\frac{1}{3}\frac{\gamma_l}{2\pi^2}\sum_{l=e,\nu_e}\int_0^{\infty}\frac{k^4}{\sqrt{k^2+m_l^2}}[f_{l}^{+}+f_l^{-}] dk+B_0 \\
\end{aligned}
\end{equation} 
 In the case of electrons, $m_e^{*}=m_e$. 
The Fermi distribution functions for particle and antiparticle are : $f_i^\pm= \frac{1}{1+exp\left({\frac{E_i(k)\mp \mu_i}{T}}\right)} $ , where $E_i(k)=\sqrt{k^2+(m_i^*)^2}$.\\
The quark number densities are given by 
\begin{equation} \label{eq:quasi_density}
\begin{aligned}
     \rho_i&=-\frac{\partial\Omega_i}{\partial\mu_i}=\frac{\gamma_i}{2\pi^2}\int_0^{\infty}k^2[f_i^{+}-f_i^{-}] dk-\\ &m_i^{*}\frac{\partial m_i^*}{\partial \mu_i}\frac{\gamma_i}{2\pi^2}\int_0^{\infty}\frac{k^2}{\sqrt{k^2+(m_i^*)^2}}[f_i^{+}+f_i^{-}] dk\\  
\end{aligned}
\end{equation}
The density of each flavor is modified due to the chemical
potential-dependent quark mass. 
The expression of pressure from Eq.~\eqref{eq:pressure_def_final} is given by
\begin{equation}
    P_{quasi}=-\Omega_{quasi}
\end{equation}
The expression of the energy density from Eq.~\eqref{eq:euler_rel}  is given by
\begin{equation} \label{eq:quasi_energy_density}
\begin{aligned}
 & \varepsilon_{quasi} =  \sum_{i=u,d,s}  \frac{\gamma_i}{2\pi^2} \int_0^{\infty} k^2 \sqrt{k^2 + (m_i^*)^2} \left( f_i^{+} + f_i^{-} \right) dk- \\
   & \sum_{i=u,d,s}\mu_i \left[m_i^{*} \frac{\partial m_i^*}{\partial \mu_i} \frac{\gamma_i}{2\pi^2} \int_0^{\infty} \frac{k^2}{\sqrt{k^2 + (m_i^*)^2}} \left(f_i^{+} + f_i^{-} \right) dk \right]  \\
   & + B_0 +  \frac{\gamma_l}{2\pi^2} \sum_{l=e,\nu_e}\int_0^{\infty} k^2 \sqrt{k^2 + m_l^2} \left( f_{l}^{+} + f_l^{-} \right) dk
\end{aligned}
\end{equation}

\subsubsection*{\textbf{ Medium effect at zero temperature}} 
The thermodynamic potential at zero temperature can be expressed as 
\begin{equation}\label{eq:quasi_pressure_zero}
\begin{aligned}
    \Omega_{T=0}=&-\sum_{f=u,d,s}\frac{1}{3}\frac{\gamma_i}{2\pi^2}\int_0^{k_i}\frac{k^4}{\sqrt{k^2+(m_f^*)^2}} dk- \\ &\frac{1}{3}\frac{\gamma_l}{2\pi^2}\sum_{l=e,\nu_e}\int_0^{k_l}\frac{k^4}{\sqrt{k^2+m_l^2}} dk+B_0 \\
\end{aligned}
\end{equation} 
where the Fermi momentum for the quark matter  $k_i=\sqrt{\mu_i^2-(m_i^*)^2}$. 
The quark number densities at zero temperature are given by
\begin{equation} \label{eq:quasi_density_zero}
\begin{aligned}
     (\rho_i)_{T=0}=&\frac{\gamma_i}{6\pi^2}\left(\sqrt{\mu_i^{*2}-m_i^{*2}}\right)^3\\ &- m_i^{*}\frac{\partial m_i^*}{\partial \mu_i^*}\frac{\gamma_i}{2\pi^2}\int_0^{k_i}\frac{k^2}{\sqrt{k^2+(m_i^*)^2}}dk \\
\end{aligned}
\end{equation}

The expression of the energy density from  Eq.~\eqref{eq:euler_rel} (with T=0) is 
\begin{equation} \label{eq:quasi_energy_density_zero}
\begin{aligned}
  \varepsilon_{quasi,T=0} = & \sum_{i=u,d,s}  \frac{\gamma_i}{2\pi^2} \int_0^{k_i} k^2 \sqrt{k^2 + (m_i^*)^2}  dk- \\
   & \sum_{i=u,d,s}\mu_i \left[m_i^{*} \frac{\partial m_i^*}{\partial \mu_i} \frac{\gamma_i}{2\pi^2} \int_0^{k_i} \frac{k^2}{\sqrt{k^2 + (m_i^*)^2}}  dk \right]  \\
   & + B_0 +  \frac{\gamma_l}{2\pi^2} \sum_{l=e,\nu_e}\int_0^{k_l} k^2 \sqrt{k^2 + m_l^2}  dk\\ 
\end{aligned}
\end{equation}
\subsubsection*{\textbf{ Medium effect with vector interactions}}  
The repulsive effect of quark interaction is also
included here by introducing the vector meson as a mediator \cite{Ju_2021,suman2023a}.
To account for the role of the vector meson, the chemical potential of the quark gets modified. The inclusion of a vector meson in the description of quark matter leads to modifications in the equation of state. The effect of the vector interaction is taken through the following relation
\begin{equation}\label{eq:vecor_int}
    \mu_i^*=\mu_i-g_VV_0
\end{equation} 
Here $g_V$ is the vector interaction coupling and $m_V$ is mass of vector mesons.
The fermi distribution function subsequently gets modified as : $f_i^{\pm}= \frac{1}{1+exp\left({\frac{E_i(k) \mp \mu_i^*}{T}}\right)} $, where $E_i(k)=\sqrt{k^2+(m_i^*)^2}$.\\
The thermodynamical potential for the quark matter along  with vector interaction reads as 
\begin{equation} \label{eq:omega_with_vector_temp}
\begin{aligned}
 \Omega_{\text{vector}}=\Omega_{\text{quasi}}-\frac{1}{2}m_V^2V_0^2
\end{aligned}
\end{equation} 
The equation of motion for the vector field  equation is given by :
\begin{equation}  
\frac{\partial \Omega_{\text{vector}}}{\partial V_0}=0
\end{equation}
\begin{align}
\implies m_V^2 V_0 = \sum_{f=u,d,s} g_{V} \rho_f
\end{align} 

\section{Results}
\label{sec:results}
In our investigation of quark matter within quark stars, we incorporate electrons into the model, necessitating consideration of chemical equilibrium conditions, charge neutrality, and baryon no conservation. In our calculations, we assign specific values to the masses of the quarks \cite{ParticleDataGroup:2020}: $m_u=2.16$ \text{MeV}, $m_d=4.67$ \text{MeV} and $m_s=93.4$ \text{MeV}. The vector interaction coupling constant is expressed as $G_V=\left(\frac{g_V}{m_V}\right)^2$ where $g_V$ and $m_V$ are from  Eq.~\eqref{eq:vecor_int} and Eq.~\eqref{eq:omega_with_vector_temp}.
\subsection{Quark matter at zero temperature} 
In the cold quark star, the microscopic conditions are expressed as :
$\mu_d=\mu_u+\mu_e=\mu_s$(chemical equilibrium), $\frac{2}{3}\rho_u-\frac{1}{3}\rho_d-\frac{1}{3}\rho_s-\rho_e=0$ (the charge neutrality condition) and $\rho=\frac{1}{3}(\rho_u+\rho_d+\rho_s)$ ( baryon number  conservation).

First, we study the thermodynamic stability condition
for the strange quark matter both with and without including the  vector interaction 
 as shown in Fig.~\ref{fig:zero_temp_eos}. The model parameters are estimated using the
stability criteria \cite{pal:23prd2_bag}. In the case  without vector interaction, we have chosen  the value of the $B_0=58~\text{MeV~fm}^{-3}$ but including the vector interaction, the value of $B_0=58 ~\text{MeV fm}^{-3}$ does not satisfy the stability criteria of Bodmer and Witten, hence a different value ($B_0=50~\text{MeV fm}^{-3}$)  was chosen. The used model parameters are
given in the figure captions. In Fig.~\ref{fig:zero_temp_eos}(a) we show the pressure vs energy density plot. The introduction of vector interaction makes the equation of state stiffer. From Fig. \ref{fig:zero_temp_eos}(b) we observe that the energy density per baryon number density reaches its minimum
value at zero pressure, ensuring thermodynamic consistency. The variation of the speed of sound with density is shown in Fig. \ref{fig:zero_temp_eos}(c). This plot reveals that the inclusion of vector interaction results in a higher speed of sound as compared to the case without  the vector interaction. Fig.~\ref{fig:zero_temp_eos}(d) illustrates the particle fraction. It is observed that the particle fractions remain more or less unchanged irrespective of the presence of vector interaction.
\subsection{Quark matter at finite temperature} 
In the case of quark matter at finite temperature, when the neutrinos are trapped in the system, the chemical equilibrium condition is modified as
\begin{equation} \label{eq:beta_eql}
\mu_d=\mu_u+\mu_e-\mu_{\nu_e}=\mu_s 
\end{equation}
Because of the neutrino trapping, the number of leptons per baryon of each flavor neutrino is conserved on a dynamical timescale :
$ Y_{Le}=Y_{e}+Y_{\nu_e} $, where $Y_{Le}$ is the total lepton fraction, $Y_e$ represents the electron fraction and $Y_{\nu_e}$ is neutrino fraction. 
    \begin{figure*}[htp]
    \centering
    \includegraphics[width=1.0\textwidth]{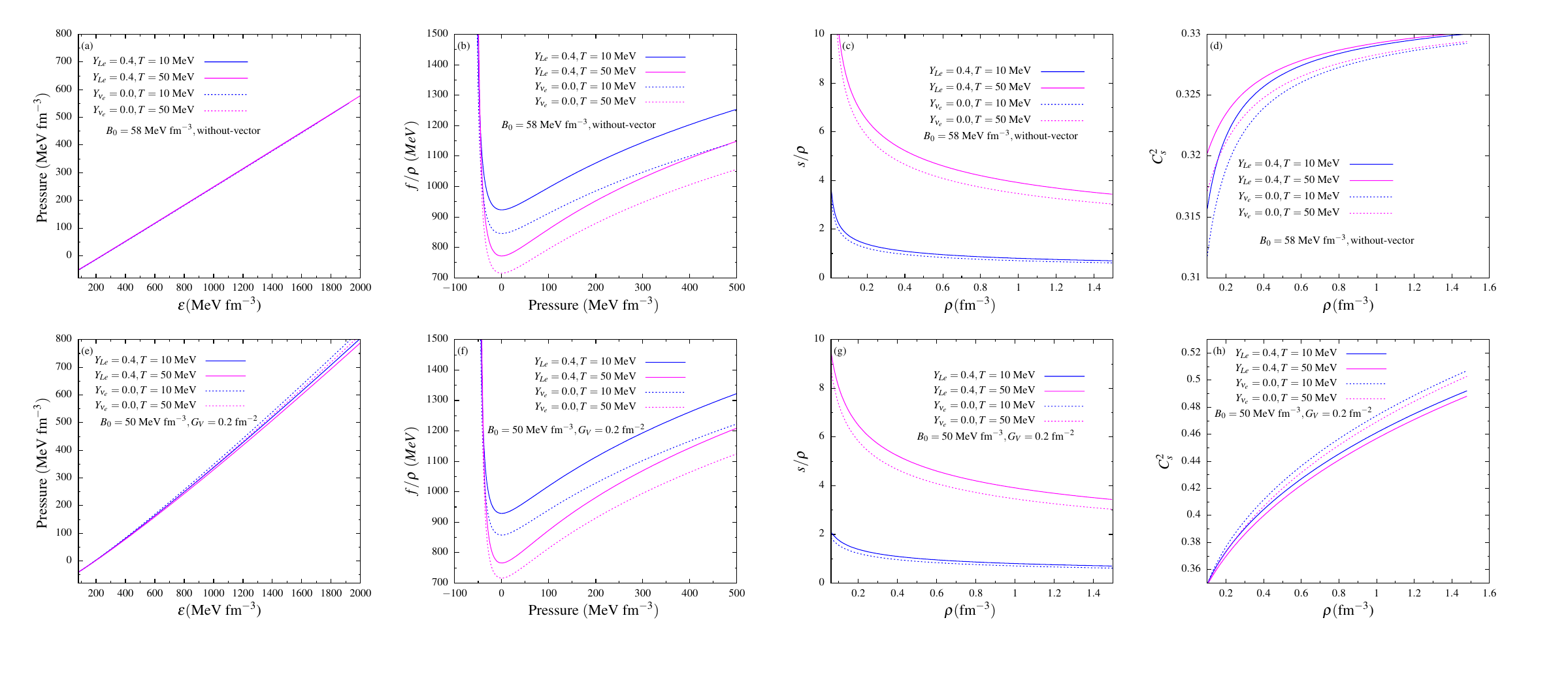}
    \caption{Plots for pressure vs. energy density, free energy density vs. pressure, entropy density, and speed of sound vs. density for temperatures $ T = 10~\text{and}~50~\text{MeV}$ , and the lepton fraction of $Y_{L_e}=0.4$ and $Y_{L_{\nu_e}}=0.0$. The upper panel shows without vector case and the lower panel shows with vector case. For the model without vector interactions (upper panel), the parameters are  $g_0 = 1.0$, $\alpha_{\mu} = 20.0$, and $B_0 = 58 \text{ MeV fm}^{-3}$. In contrast, for the model with vector interactions (lower panel), the parameters are $g_0 = 1.0$, $\alpha_{\mu} = 20.0$, $B_0 = 50 \text{ MeV fm}^{-3}$, and $G_V = 0.2 \text{ fm}^{-2}$.}  
    \label{fig:T_input_eos}
   \end{figure*}  
After the formation of a proto-quark star, the key parameters influencing the state of matter in its hot interior include the baryon density, electron fraction, and neutrino fraction. Detailed information about the proto neutron star can be found in \cite{Prakash:1996xs,Shao:2011nu,Steiner:2001rp,Chu:2022ofc,Lopes:2020dvs,Goussard:1996dp,Sandin:2007zr,Chu:2017huf,Chu:2021aaz,Chu:2019ipr,Chu:2018dch}. We examine the constituents of matter under two scenarios: neutrino trapped and neutrino transparent. In the neutrino-trapped scenario, the composition of matter is determined by chemical equilibrium and a fixed lepton fraction. For our study, we use a lepton fraction value of $Y_{Le}=0.4$. In the neutrino-transparent scenario$(Y_{\nu_e}=0)$, the composition of the matter is calculated from the chemical equilibrium condition. 
In this work, we consider both isothermal and isentropic processes, the details of which are outlined below.

\begin{figure*}[htp] 
    \centering
    \includegraphics[width=0.90\textwidth]{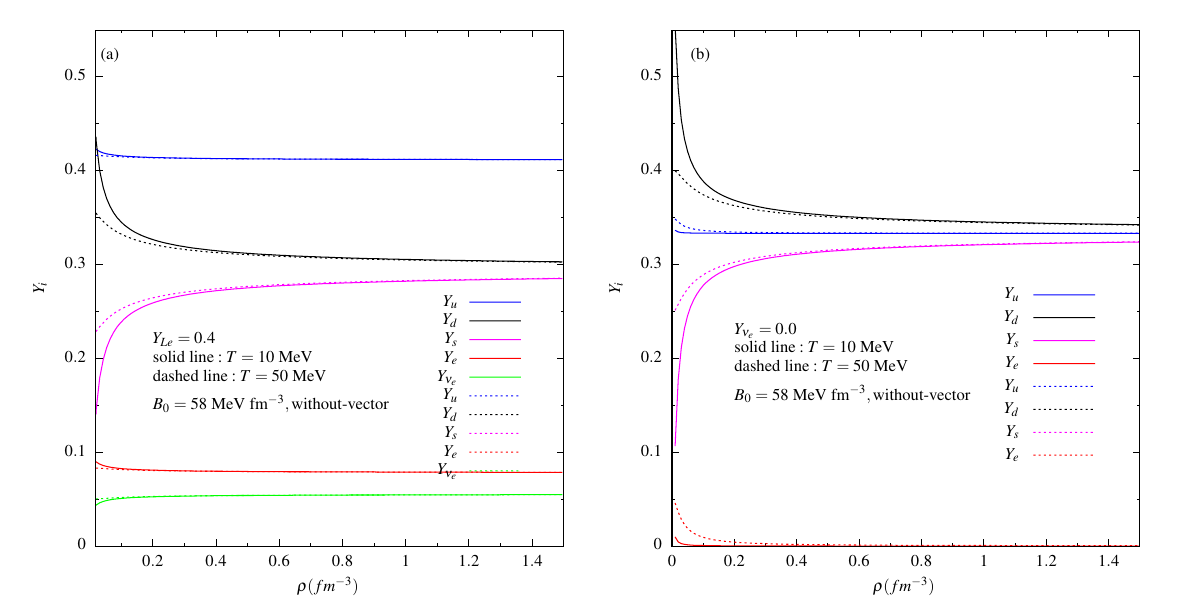}
    \caption{The particle fraction ($Y_i=\frac{\rho_i}{3\rho}$ for i=u, d and s quarks  and for leptons($l=e,\nu_e$)  $Y_l=\frac{\rho_l}{\rho}$ ) plots for (a) neutrino trapped (b) neutrino transparent for two different temperatures. }   
    \label{fig:T_input_partl_frac}
\end{figure*}  

\begin{figure*}[htp]
    \centering
    \includegraphics[width=1.0\textwidth]{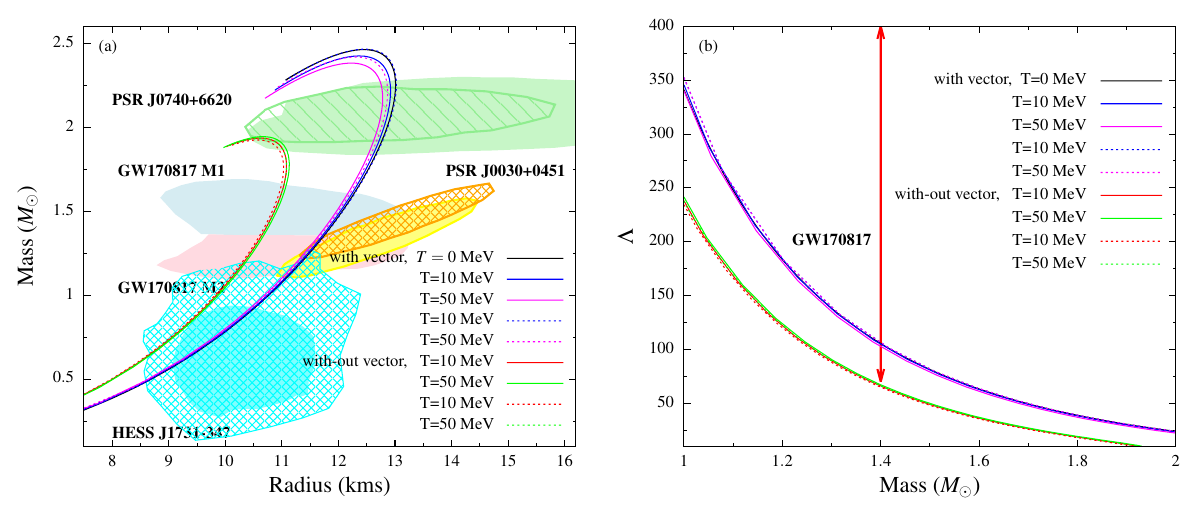}
    \caption{(a) Mass-radius relationship of quark star configuration for the EoSs obtained using quasi-particle model, the model parameters with vector interactions are $g_0 = 1.0$, $\alpha_{\mu} = 20.0$, $B_0 = 50 \text{ MeV fm}^{-3}$, and $G_V = 0.2 \text{ fm}^{-2}$ for different values of temperature and lepton fractions and the model parameters without vector interactions are $g_0 = 1.0$, $\alpha_{\mu} = 20.0$, and $B_0 = 58 \text{ MeV fm}^{-3}$.  Observational limits imposed from PSR J0740+6620\cite{Fonseca:2021wxt}, PSR J0030+0451 \cite{Riley:2019yda,Miller:2019cac} and HESS J1731-347\cite{2022NatAs}are indicated. The constraints on the $M-R$ plane prescribed from GW170817\cite{LIGOScientific:2018cki}  are also compared. (b) Variation of tidal deformability with mass for different values of above-mentioned parameters. The constraints on $70\le\Lambda_{1.4}\le 580$\cite{LIGOScientific:2018cki} are included.  The solid lines represent the neutrino trapped matter ($Y_{Le}=0.4$) and the dotted lines represent the neutrino transparent matter ($Y_{\nu_e}=0$) }
    \label{fig:T_input_MR}
\end{figure*} 
\begin{figure*}[htp]
    \centering
    \includegraphics[width=1.0\textwidth]{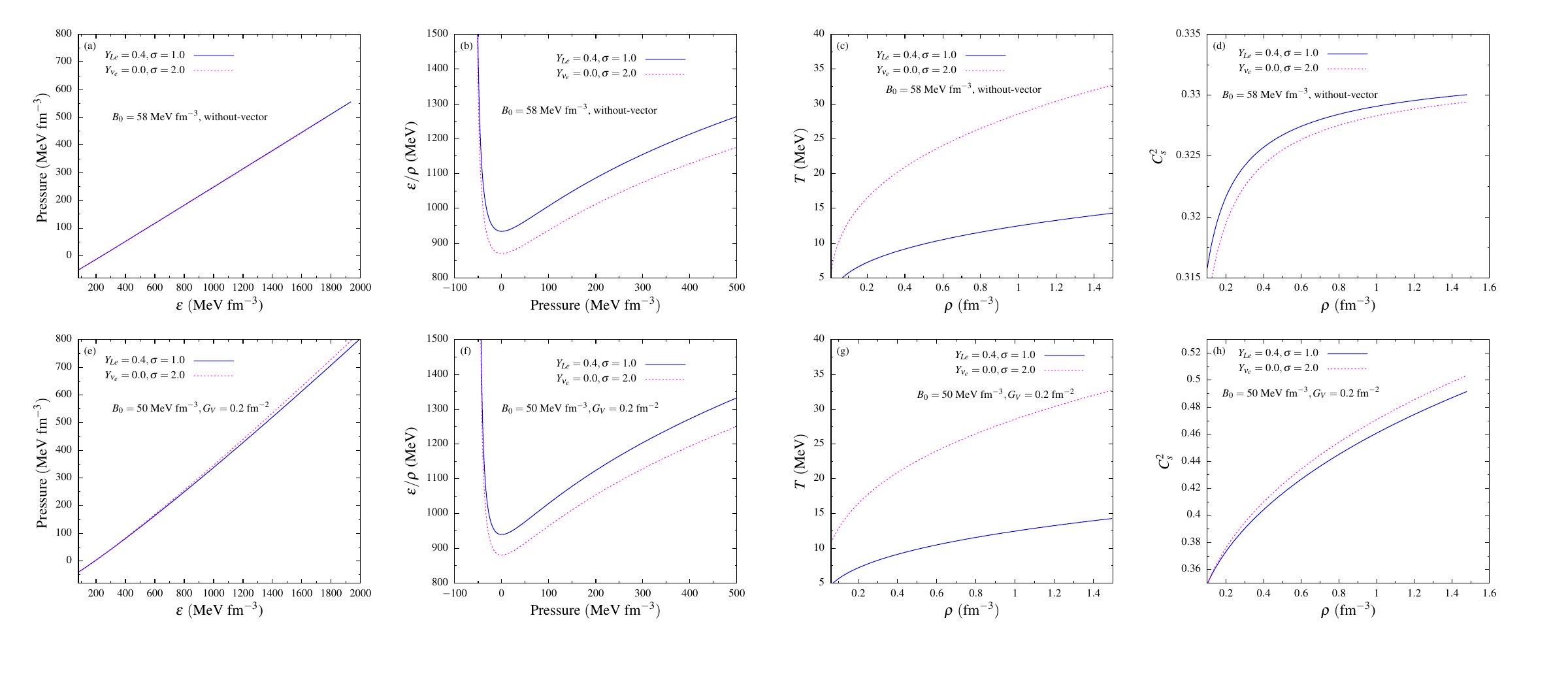}
    \caption{Plots for pressure vs. energy density, free energy density vs. pressure, entropy density, and speed of sound vs. density for $ \sigma = 1,Y_{Le}=0.4~\text{and}~\sigma=2.0,Y_{\nu_e}=0.4$. The upper panel shows without vector case and the lower panel shows with vector case. For the model without vector interactions (upper panel), the parameters are  $g_0 = 1.0$, $\alpha_{\mu} = 20.0$, and $B_0 = 58 \text{ MeV fm}^{-3}$. In contrast, for the model with vector interactions (lower panel), the parameters are $g_0 = 1.0$, $\alpha_{\mu} = 20.0$, $B_0 = 50 \text{ MeV fm}^{-3}$, and $G_V = 0.2 \text{ fm}^{-2}$.}  
    \label{fig:sigma_input_eos}
\end{figure*}  
\begin{figure}[htbp] 
	\centering
	\includegraphics[width=0.45\textwidth]{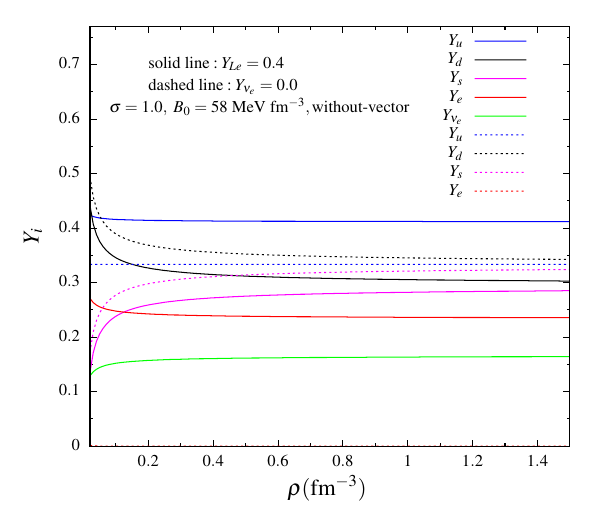}
	\caption{Particle fraction plots for 
 constant entropy desnity per baryon desnity and lepton fraction. }   
	\label{fig:sigma_input_partl_frac}
\end{figure}
\subsection*{Quark matter at fixed temperature}  
The description of the PQS in the isothermal process is determined for a given density with different temperatures and lepton fractions. The equation of state is characterized by
\begin{equation} 
\begin{aligned}
    X=&X(\rho,T,Y_{Le})~~\textbf{for trapped neutrino}\\
    X=&X(\rho,T)~~\textbf{for transparent neutrino}\\
\end{aligned}
\end{equation} 
where $X$ is the equation of state.
In this description, the entropy density per baryon density $(\sigma)$ is varied with the baryon density.
As we have mentioned earlier in the introduction section, our main aim in this work is to provide a self-consistent thermodynamic treatment in the quasi-particle model for the description of the quark matter in the context of the quark star.
In Fig.~\ref{fig:T_input_eos}, we examine the thermodynamic properties of strange quark matter within the framework of the PQS, comparing scenarios with and without vector interactions. The model parameters are estimated using the stability criteria \cite{torres2013quark}. The used model
parameters are given in the figure caption. The upper panel represents cases without vector interaction. In the trapped neutrino case with a lepton fraction $Y_{Le}=0.4$, we have taken two temperatures $T=10$ and $50$ \text{MeV}.  The main result from the thermodynamical consistency point of view is that the value of free energy density must vanish at zero pressure and also we calculate the equation of state respecting the Euler relation. From Fig.~\ref{fig:T_input_eos}(a), we find the equation of state is almost insensitive to the temperature and lepton fraction. In Fig.~\ref{fig:T_input_eos}(b), we show the thermodynamic stability for the above-mentioned cases; as the temperature $(T)$ increases, the minimum values of the free energy density per baryon density get lower, and also in the neutrino transparent case, the minimum of free energy density per baryon density gets lower.  Next, we plot entropy density per baryon density( $\sigma$) in Fig.~\ref{fig:T_input_eos}(c). As the density increases, the value of $\sigma$ decreases and then nearly saturates to fixed values. The value of the $\sigma$ in the neutrino transparent case is lower as compared to the trapped neutrino case for a fixed temperature. 
In Fig.~\ref{fig:T_input_eos}(d), we show the variation of the speed of sound with different temperatures and lepton fractions. The value of $C_s^2$ is higher at higher temperatures and neutrino transparent cases.
In the lower panel of Fig.~\ref{fig:T_input_eos}, we study the above-mentioned observables with vector interaction. All the thermodynamic properties are consistent as mentioned above.
The inclusion of vector interaction significantly impacts the equation of state, leading to notable changes in its values across different temperatures and lepton fractions. The EoS becomes more sensitive to temperature and lepton fraction when the vector interaction is taken into account as seen from Fig.~\ref{fig:T_input_eos}(e). The results for $\frac{f}{\rho}$ and $\frac{s}{\rho}$ are almost similar irrespective of the addition of the vector term as seen from the comparison of the upper and lower panel figures. The pattern of variation as well as the value of $C_{s}^{2}$ changes with the inclusion of vector interaction as seen from  Figs.~\ref{fig:T_input_eos}(d) and ~\ref{fig:T_input_eos}(h). 
In the case  without vector interaction (Fig.~\ref{fig:T_input_eos}(d)), at the lower density it rises fast and at higher density it saturates. In the vector interaction case (Fig.~\ref{fig:T_input_eos}(h)), $C_s^2$ increases monotonically with density. The inclusion of the vector interactions increases the magnitude of $C_s^2$ compared to the case without vector interactions. The composition of the PQS changes significantly with evolving temperature and lepton fraction.  For the study of the proto quark star, we consider one lower temperature (T=10 MeV) and another higher temperature (T=50 MeV), both for neutrino-trapped and neutrino-free matter. Here, the vector interaction case is not included because the particle fraction remains nearly the same as in the case without vector interactions. 
 We show the particle fraction plot in Fig.~\ref{fig:T_input_partl_frac}.  In Fig.~\ref{fig:T_input_partl_frac}(a), we study the particle fraction for the neutrino trapped matter for $T=10~\text{MeV}$ and $T=50~\text{MeV}$. In Fig.~\ref{fig:T_input_partl_frac}(b), we study the particle fraction for the neutrino free matter for $T=10~\text{MeV}$ and $T=50~\text{MeV}$. At lower densities, particle fractions differ slightly for the two temperatures whereas there is almost no difference as one increases the density both for neutrino trapped matter and neutrino transparent matter. The difference in results at lower densities for the two temperatures is more for the d and the s quarks as compared to others. The fraction of u quarks depends on the electron fraction ($ Y_u=\frac{1}{3}+\frac{Y_e}{3}$), which does not vary significantly with temperature; therefore, the temperature has a minimal effect on u quark fraction. On the other hand, the fractions of d and s quarks are related by using charge neutrality and baryon number conservation as $Y_d+Y_s=\frac{2}{3}-\frac{ Y_e}{3}$, implying that an increase in one fraction necessitates a decrease in the other ($Y_e$ changes very little). Since the mass of the strange quark is  considerably greater than that of the d quark, an increase in temperature leads to an increase in $Y_s$ and a corresponding 
decrease in $Y_d$. From this figure, we find that there is a change in the particle fraction plot in the two cases due to the involvement of neutrino. The fraction of the u quark is more for the neutrino trapped matter (Fig.~\ref{fig:T_input_partl_frac}(a)) as compared to the neutrino transparent(Fig.~\ref{fig:T_input_partl_frac}(b)) case due to the $\beta$ equilibrium reaction Eq.~\eqref{eq:beta_eql}.

In Fig.~\ref{fig:T_input_MR}(a), we show the results of the M-R relation using the PQS equation of state for the isothermal cases with vector interactions and without vector interactions. The model parameters utilized are detailed in the figure caption of Fig.~\ref{fig:T_input_MR}.
In the case  without vector interaction, 2$M_{\odot}$ constraints is just satisfied, but PSR J0030+0451 constraints is not satisfied. In the case of vector interaction,
the PQS configurations comply with various astrophysical constraints from different observations, including the measurements of PSRJ0740+6620 and PSRJ00030+0451 from the NICER experiment, as well as data from GW170817 and HESS~J1731-347. The results with vector interaction are more sensitive to temperature as well as trapped neutrinos than the case without including the vector interaction.  
In Fig.~\ref{fig:T_input_MR}(b), we examine the variation of the tidal deformability with mass.
The tidal deformability constraints from $\text{GW170817}$ are met when the vector interaction is included. However, without the vector interaction, these constraints are not satisfied. Thus, incorporating the vector interaction ensures compliance with the astrophysical constraints. 
\subsection*{Quark matter at fixed entropy density per baryon density}
The description of the PQS in the isentropic process is determined for a given density with different entropy densities per baryon density
 and lepton fraction. The equation of state is characterized by
\begin{equation} 
\begin{aligned}
    Y=&Y(\rho,\sigma,Y_{Le})~~\textbf{for trapped neutrino}\\
    Y=&Y(\rho,\sigma)~~\textbf{for transparent neutrino}\\
\end{aligned}
\end{equation}  
In this description, temperature $(T)$ varies with density.
As we have mentioned in the earlier section also,  our main aim in this work is to study the self-consistent thermodynamics of quark matter in different environments. In this environment, we consider three different snapshots of the isentropic process, which has been thoroughly explored in earlier studies\cite{Prakash:1996xs,Shao:2011nu,Steiner:2001rp,Chu:2022ofc,Lopes:2020dvs}. At the first snapshot of the PQS, after the gravitational collapse with the explosion of a supernova 
$\sigma=1,~Y_{Le}=0.4$. In the following tens of seconds, the diffusing neutrinos heat the star and escape from the star which could increase the corresponding entropy density, and then the decreasing neutrino fraction reaches almost zero i.e, the condition is 
$\sigma=2,~Y_{\nu_e}=0$.
After the heating stage, the stars begin to cool down and eventually transform into cold quark stars ($\sigma=0,~Y_{\nu_e}=0$).

In Fig.~\ref{fig:sigma_input_eos}, we study the equation of state at two entropy densities and lepton fractions ($\sigma=1.0, Y_{Le}=0.4;\sigma=2.0, Y_{\nu_e}=0.0$). The upper panel shows the equation of state without vector interaction and the lower panel displays that with vector interaction.  
\begin{figure*}[htp]
    \centering
    \includegraphics[width=1.0\textwidth]{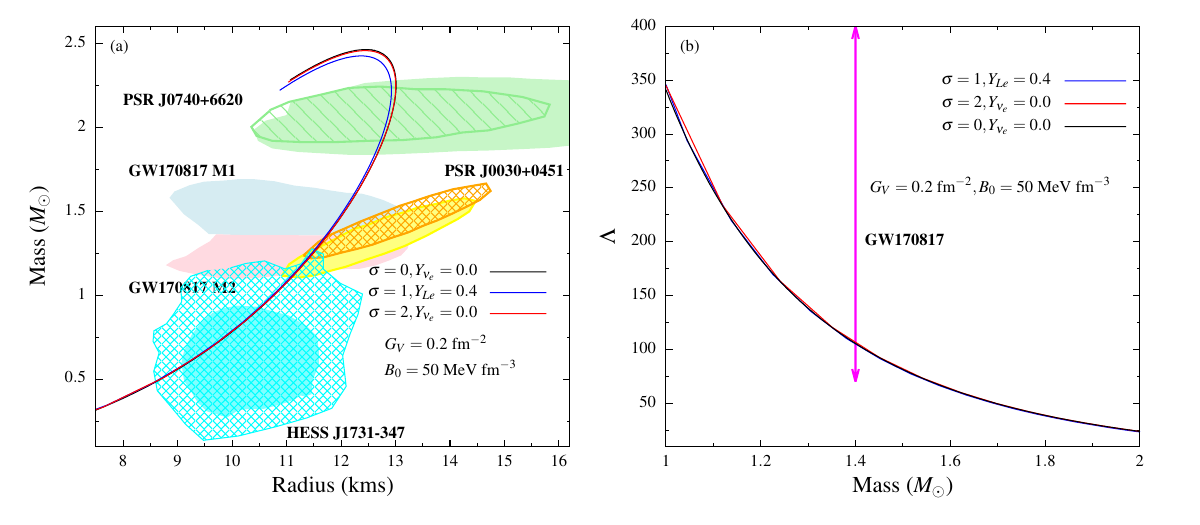}
    \caption{(a)Mass-radius relationship of quark star configuration for the EoSs obtained using quasi-particle model, the model parameters are: $g_0 = 1.0$, $\alpha_{\mu} = 20.0$, $B_0 = 50 \text{ MeV fm}^{-3}$, and $G_V = 0.2 \text{ fm}^{-2}$ for different values of entropy density per baryon density and lepton fractions. Observational limits imposed from PSR J0740+6620\cite{Fonseca:2021wxt},PSR J0030+0451 \cite{Riley:2019yda,Miller:2019cac} and HESS J1731-347\cite{2022NatAs}are indicated. The constraints on the $M-R$ plane prescribed from GW170817\cite{LIGOScientific:2018cki}  are also compared. (b)Variation of tidal deformability with mass for different values of above mention parameter. The constraints on $70\le\Lambda_{1.4}\le 580$\cite{LIGOScientific:2018cki} are also included.}   
    \label{fig:sigma_input_MR}
\end{figure*} 
As we have already mentioned in the introduction section,  we need to consider the energy density and not the free energy density for an isentropic process for the thermodynamical consistency i.e, the value of energy density must vanish at zero pressure, and also one needs to calculate the equation of state respecting the Euler relation.

In Fig.~\ref{fig:sigma_input_eos}(b), we present the thermodynamic stability for the scenarios mentioned above. As  $\sigma$ increases, the minimum values of the $(\varepsilon/ \rho)$ get lower. Next, we plot the temperature profile in the Fig.~\ref{fig:sigma_input_eos}(c). As the density increases, the temperature increases. In Fig.~\ref{fig:sigma_input_eos}(d), we show the variation of the speed of sound with density for above mentioned parameters. The introduction of vector interaction is more sensitive to the temperature as seen from Fig.~\ref{fig:sigma_input_eos}(e). In Fig.~\ref{fig:sigma_input_eos}(f), we study the thermodynamical stability, and in Fig.~\ref{fig:sigma_input_eos}(g), we study the temperature profile with the density in presence of vector interaction. As discussed in the case of isothermal scenarios, the inclusion of the vector interaction modifies the pattern of the speed of sound, as shown in Fig. Fig.~\ref{fig:sigma_input_eos}(h).

As mentioned in the earlier section, the composition of the PQS changes with evolving temperature and lepton fraction.
We show the particle fraction plot in the isentropic process for neutrino-trapped  and neutrino-free case with $(\sigma=1)$ shown in the Fig.~\ref{fig:sigma_input_partl_frac}.  From this figure, we find that there is a change in the particle fraction in the two cases due to the inclusion of neutrino. 
 The objects born in the core collapse supernovae are extremly hot, the temperature range being of the order of 1-100 MeV. In this work we consider the situation of the proto quark star for wich $\sigma$ lies in between $0-2$. The temperature corresponding to the value of the  $\sigma=1$ is  nearly 15 MeV and that for $\sigma=2$  it is  nearly 35 MeV (as seen from the Fig.~\ref{fig:sigma_input_eos}(c)).
In Fig.~\ref{fig:sigma_input_MR}(a), we show the results of the M-R relation using the PQS equation of state with vector interaction for the isentropic cases. We include the constraints from $\text{PSRJ0740+6620}$, PSRJ0030+0451,$\text{GW170817}$, and $\text{HESS~J1731-347}$. As anticipated from the equation of state, the thermal effect has a minimal impact on the mass-radius diagram. The impact of the neutrino lepton fraction on the mass-radius diagram is also minimal.  
In the  Fig.~\ref{fig:sigma_input_MR}(b), we study the variation of tidal deformability with mass and it is observed that the tidal deformablity constraints from the $\text{GW170817}$ are satisfied. 

\section{SUMMARY AND CONCLUSION}\label{sec:conclusion}
In this work, we have studied the quasi-particle model where the medium effect has been incorporated via chemical potential-dependent quark mass. We work in the grand canonical ensemble ensuring  proper choice of thermodynamical variables\cite{pal:23prd2_bag}. Unlike previous versions of the model \cite{Chu:2021aaz, Schertler:1996tq}, where an external counterterm was introduced to the thermodynamic potential to cancel the additional  contribution to the density term,  in this work, we begin with the grand-canonical partition function of a free Fermi gas with a chemical potential-dependent quark mass. Our method guarantees thermodynamic consistency without requiring any additional counterterms in the thermodynamic potential. We carry out a detailed self-consistent thermodynamic analysis for both cold and thermal equations of state. In the case of the finite temperature EoS, we study both the isothermal as well as the isentropic processes. Apart from  ensuring the self-consistent thermodynamical formulation,  we also study the structural properties, namely mass-radius and dimensionless tidal deformability  of quark stars using the quasi-particle model.  
The inclusion of the vector interaction in the quasi-particle model ensures fulfillment of the recent astrophysical observational constraints.

\bibliography{bagqm}

\end{document}